\documentclass{iaus}
\usepackage{graphicx}

\title[Dynamic Models of the Sun] 
{Dynamic Models of the Sun from the Convection Zone to the Chromosphere}

\author[S. Wedemeyer-B\"ohm]   
{Sven Wedemeyer-B\"ohm$^1$}

\affiliation{$^1$Kiepenheuer-Institut f\"ur Sonnenphysik, \break
Sch\"oneckstr. 6, 79104 Freiburg, Germany \break email: wedemeyer@kis.uni-freiburg.de\\[\affilskip]
}

\volume{239}  
\pubyear{2007}
\pagerange{119--126}                  
\date{?? and in revised form ??}
\jname{Convection in Astrophysics}
\editors{F. Kupka, I.W. Roxburgh \& K.L. Chan, eds.}

\begin{document}

\maketitle

\begin{abstract}
  The chromosphere in internetwork regions of the quiet Sun was
  regarded as a static and homogeneous layer for a long time. Thanks
  to advances in observations and numerical modelling, the wave nature
  of these atmospheric regions received increasing attention during
  the last decade. Recent three-dimensional radiation
  magnetohydrodynamic simulations with CO5BOLD feature the
  chromosphere of internetwork regions as a dynamic and intermittent
  phenomenon. It is a direct product of interacting waves that form a
  mesh-like pattern of hot shock fronts and cool post-shock regions.
  The waves are excited self-consistently at the top of the convection
  zone. In the middle chromosphere above an average height of 1000 km,
  plasma beta gets larger than one and magnetic fields become more
  important. The model chromosphere exhibits a magnetic field that is
  much more homogeneous than in the layers below and evolves much
  faster. That includes fast propagating (MHD) waves. Further
  improvements of the simulations like time-dependent hydrogen
  ionisation are currently in progress. This class of models is
  capable of explaining apparently contradicting diagnostics such as
  carbon monoxide and UV emission at the same time.
\end{abstract}

\firstsection 
              
\section{Introduction}

The atmosphere of the Sun is a very dynamic and inhomogeneous
layer that exhibits phenomena on a large range of spatial and temporal
scales.
A self-consistent and comprehensive numerical model that can match all
accessible chromospheric diagnostics would be of great value. 
The complexity of the object,
however, so far limited all efforts to more or less crude 
approximations or at least to addressing certain aspects only.
Modelling of these layers is numerically very demanding since a
realistic description requires the inclusion of many physical
ingredients.  A three-dimensional and time-dependent modelling is mandatory 
\cite[(cf. Carlsson \& Stein 1995; Skartlien \etal\ 2000; Stein \& Nordlund 1989, 2006; and many more)]{carlsson95, skartlien00c, stein89, 2006ApJ...642.1246S}. 
In order to make progress, the radiation magnetohydrodynamics code
\mbox{\textsf{CO}$^\mathsf{5}$\textsf{BOLD}}
\cite[(Freytag \etal\ 2002)]{cobold} 
has been upgraded recently.  It now offers a large range of possible
applications, including the time-dependent treatment of chemical 
reaction networks, hydrogen ionisation, and magnetic fields.

This paper addresses the potential of the currently possible
radiation magnetohydrodynamics simulations and the 
synthesis of intensity maps for interpreting observations and even
defining constraints for future observation campaigns for internetwork
regions of the Sun.  To demonstrate this, a new three-dimensional model is
used as input for the calculation of synthetic intensity images.

\section{Time-dependent radiation (magneto-)hydrodynamics}\label{sec:code}

The numerical simulations presented here were all carried out with 
an upgraded version of 
\mbox{\textsf{CO}$^\mathsf{5}$\textsf{BOLD}} 
\cite[(Freytag \etal\ 2002, Wedemeyer \etal\ 2004, Schaffenberger \etal\ 2005)]{cobold,wedemeyer04a,schaffenberger05},   
a radiation magnetohydrodynamics code for simulating the convective
layers and atmospheres of stars.  Additional features are the
time-dependent treatments of dust formation
\cite[(H\"ofner \etal\ 2004)]{hoefner04},  
chemical reaction networks 
(\cite[Wedemeyer-B\"ohm \etal\ 2005, 2006]{wedemeyer05a,wedemeyer06a}),  
and also non-equilibrium hydrogen ionisation 
(\cite[Leenaarts \& Wedemeyer-B\"ohm 2006a, 2006b, hereafter L06]{leenaarts06a, leenaarts06b}).
Radiative cooling due to spectral lines of carbon monoxide
can now be taken into account 
\cite[(Wedemeyer-B\"ohm \& Steffen submitted, hereafter W06)]{wedemeyer06a}.
Operator splitting allows to treat hydrodynamics, tensor viscosity,
chemistry, non-equilibrium hydrogen ionisation, and radiative transfer
in subsequent steps of this order.  The overall computational time
step is typically around 0.1\,s to 0.2\,s for non-magnetic simulations
but up to a factor ten or more smaller for the MHD case, depending on
magnetic field strength.  The lateral boundary conditions are periodic
whereas the lower boundary is ``open'', i.e. material can flow in and
out of the computational box. The upper boundary condition is usually
of 'transmitting' type so that shock waves can leave the computational
domain without being reflected.

\section{The numerical model}
\label{sec:model}

\begin{figure}
  \begin{center}
  \includegraphics[width=12cm]{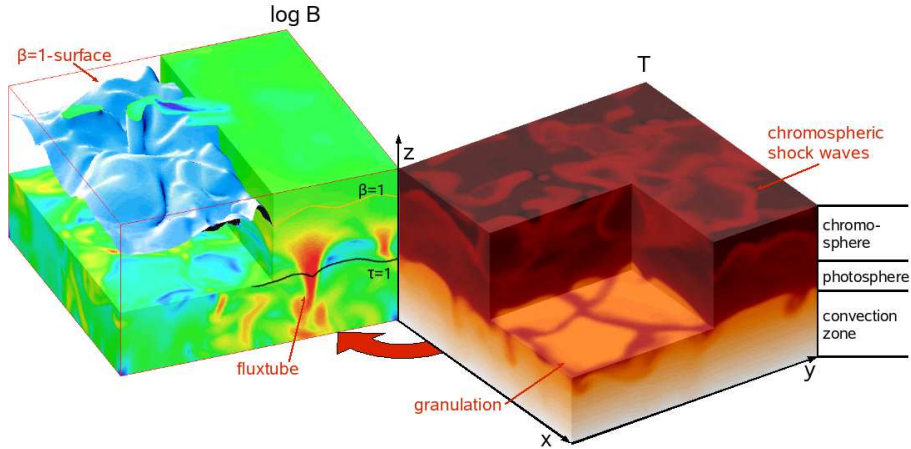}
  \end{center}
  \caption{Three-dimensional magnetohydrodynamics model: Gas temperature $T$  
  (right cube) and logarithmic magnetic field strength $\log B$ (cube flipped open to the left).    
  }
 \label{fig:cube}
\end{figure}

The new 3D model  presented here (see Fig.~\ref{fig:cube})
is similar to the model by 
\cite[Wedemeyer \etal\ (2004, hereafter W04)]{wedemeyer04a}
and
\cite[Schaffenberger \etal\ (2005, hereafter S05)]{schaffenberger05} 
but was calculated with non-grey radiative transfer (multi-group with ATLAS6 opacities, 
\cite[Kurucz 1970]{kurucz70}) 
and time-dependent CO chemistry with the same chemical reaction network as 
used by
\cite{wedemeyer05a}.
With a horizontal extent of $4800$\,km $\times$ 4800\,km 
(corresponding to $\approx 6''.6\,\times\,6''.6$ in ground-based
observations) it is slightly smaller than the model by W04.   
It reaches from \mbox{$z = -1400$\,km} in the
convection zone to the middle chromosphere at a height of 
$z \sim 1700$\,km.
The origin of the geometric height scale (\mbox{$z =
 0$~km}) is adjusted to the average Rosseland optical depth unity.
The photosphere is here referred to as the layer between $0$~km and
$500$~km in model coordinates, and the chromosphere as the layer
above.  The spatial resolution is $\Delta x = \Delta y = 40$\,km in
horizontal directions, whereas the grid cells are non-equidistantly
spaced in vertical direction ($\Delta z = 46$~km at the bottom,
decreasing with height to $12$~km for $z \ge -270$\,km).  
The treatment of radiative transfer is switched to grey 
for the chromosphere. The initial field has a strength of 
$B_0 = 10$\,G only and is vertical and unipolar.

\section{Structure and dynamics of the model atmosphere}

The new model is very similar to recent other models of the 
\mbox{\textsf{CO}$^\mathsf{5}$\textsf{BOLD}} group concerning 
atmospheric structure and dynamics of solar internetwork regions
(see W04, S05, L06 etc.).  
Altogether the simulations yield a new picture of the solar atmosphere in
internetwork regions as highly structured, intermittent, and dynamic
phenomenon.  A key role play the ubiquitous propagating and
interacting shock waves and adiabatically expanding post-shock regions
that produce an intermittent mix of co-existing hot and cool regions
in the chromosphere.  But also the photosphere below
exhibits structure on very small scales.  
The reversed granulation pattern in the middle photosphere provides
cool regions where the largest absolute amount of carbon monoxide
molecules is located.  The panels a and b of Fig.~\ref{fig:int} show
gas temperature and CO number density in the middle photosphere at $z
= 210$\,km, respectively.  Gas temperature and CO number density are
closely anti-correlated (-0.86), owing to the strong temperature sensitivity
of carbon monoxide. The cool areas in the middle photosphere allow for
formation of more CO compared to the hotter ``filamentary'' regions by
typically 1-2 orders of magnitude.  But CO is also very abundant in
the (low) chromosphere above and binds a large fraction of all
available carbon atoms there.  An exception are shocks where CO is
significantly depleted by dissociation.  The distribution of CO with a
peak in absolute number density on average at $z = 150$\,km is in line with
the findings by
\cite[Uitenbroek (2000a,b)]{uitenbroek00a, uitenbroek00b}
and
\cite{asensio03}. 
The shock waves also hamper the cooling action of infrared CO lines.
Due to the continuous passage of these waves the atmosphere cannot
relax to a cool state (W06). The average temperature is only
reduced by $\sim 100$\,K. 

In sharp contrast to the thermal structure of the model chromosphere,
the ionisation degree of hydrogen is rather homogeneous in the upper
layers at values set by the passing hot wave fronts (L06).  A similar
picture is found for the magnetic field in the chromosphere of
internetwork regions that is much more homogeneous and also more
dynamic than in the layers below (S05).  The highly dynamic middle
chromosphere seems to be separated from the slower evolving lower layers
by the surface of plasma $\beta = 1$ (at an average height of $z =
$1000\,km).  The photospheric field is highly structured with a kind
of ``small-scale canopy'' above granule interiors as a result of flux
expulsion by the hydrodynamic flow. Even a weak initial field of
10\,G is sufficient to build up kG flux tubes whereas the field
strength above the granule interiors often drops to only $\sim 3$\,G.
This finding is still compatible with observations by 
\cite{2004Natur.430..326T}
using the Hanle effect (see S05).  

\section{Synthetic intensity maps}

\begin{figure}
 \includegraphics{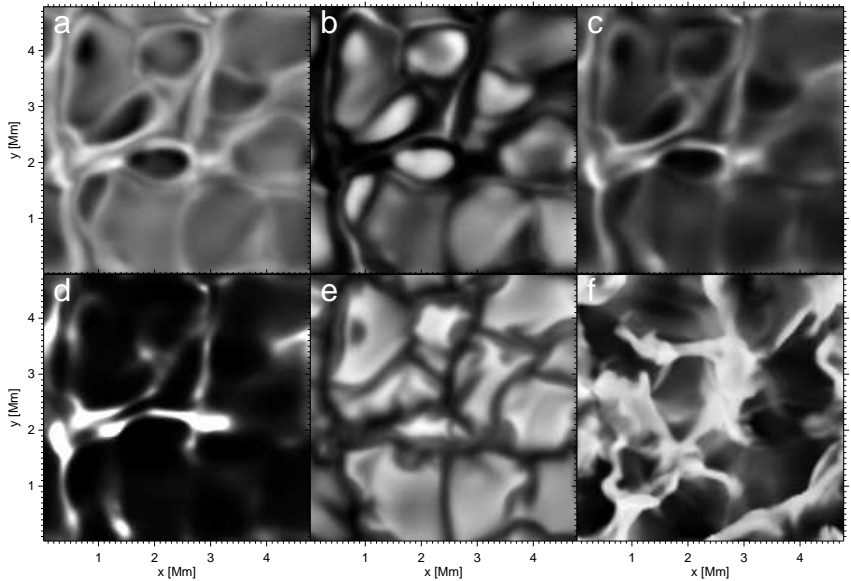}
 \caption{Non-grey three-dimensional model: 
   (\textit{a}) Gas temperature at a height of $z = 220$\,km, 
   (\textit{b}) CO number density at the same height, 
   (\textit{c}) emergent intensity at 396.7\,nm in the line wing of Ca\,II\,H, 
   and in the continuum at 
   (\textit{d}) 160\,nm, 
   (\textit{e}) 500\,nm, and
   (\textit{f}) 1\,mm (grey RT). 
   The data range of 
   panel~d is limited to 30\,\% of its maximum value.
 }\label{fig:int}
\end{figure}

A snapshot from the 3D simulation described above 
is used as input for radiative transfer calculations 
with 
\textsf{Linfor3D} ({\texttt{http://www.aip.de/$\sim$mst/linfor3D\_main.html}}) 
for a small selection of spectral features: The line wing of Ca\,II\,H at 369.7,nm
and the continuum at wavelengths of 160\,nm, 500\,nm, and 1\,mm (see Fig.~\ref{fig:int}).  
All intensity maps refer to disk-centre.

The continuum intensity at a wavelength of 500\,nm
(Fig.~\ref{fig:int}e) clearly exhibits the granulation pattern in an 
internetwork region (i.e.  with weak magnetic field only).  It has an
intensity contrast of $\delta I_\mathrm{rms} / <I>\, = 22$\,\%,
whereas the corresponding contrast of the emergent grey intensity is
17\,\% and thus the same as for the 3D hydrodynamic model with grey
radiative transfer by W04.

The Ca\,II\,H line wing intensity 
\cite[(Fig.~\ref{fig:int}c, see also Leenaarts \& Wedemeyer-B\"ohm 2005)]{leenaarts05} 
originates from the middle photosphere and thus shows the same although reversed
pattern as the CO distribution (panel b).  The contrast is 41\,\% and
thus much higher than for the 500\,nm continuum.  The intensity at
160\,nm is on average formed only a little bit higher in the high
photosphere.  In contrast to the Ca\,II\,H line wing map, the intensity
image at 160\,nm only exhibits the hottest regions of the reversed
granulation pattern as the gas temperature translates highly
non-linearly into intensity at UV wavelengths.  Most of the pattern
appears relatively dark and is thus hard to detect so that the hot
regions resemble isolated grainy features.  The contrast is quite
extreme and reaches 190\,\% for the displayed map.  
Already the resolution of 0.5'' (or more) of the 160\,nm channel of the 
Transition Region and Coronal Explorer 
\cite[(TRACE, Handy \etal\ 1999; Schneider \etal\ 2004)]{1999SoPh..187..229H,2004Icar..168..249S}
might be too low for detecting the faint background pattern. 
In combination
with CO and Ca line wing diagnostics, the UV continuum is nevertheless
a valuable atmospheric probe.  The middle photosphere can also be
observed in the cores of spectral lines as, e.g., Fe\,I at 709.04\,nm
\cite[(Janssen \& Cauzzi 2006)]{janssen06}.

The situation is more complicated for the solar chromosphere as (i)
meaningful diagnostics are rare and/or hard to interpret and (ii)
realistic multi-dimensional numerical modelling of this layer is
difficult due to many reasons and hardly possible with nowadays
computational resources.  The central parts of the Ca\,II resonance
lines are commonly used as chromospheric diagnostic
\cite[(see W\"oger \etal\ 2006 for a recent example)]{woeger06}
but are difficult to interpret as the intensity has contributions from
a large height range and non-local / non-equilibrium effects have to
be taken into account.  In contrast, the gas temperature almost
linearly translates into continuum intensity in the (sub-)millimetre
range (see Fig.~\ref{fig:int}f for a continuum intensity map at
1\,mm).  Unfortunately the current instruments are not able to resolve
the small spatial scales discussed here
\cite[(see, e.g., White \etal\ 2006)]{white06}.
The future Atacama Large Millimeter Array 
\cite[(ALMA, Beasley \etal\ 2006)]{beasley06}, 
a interferometer with a large number of antennae in the
(sub-)millimetre range, will provide sufficiently high spatial
resolution for a detailed comparison with numerical models.  The
receivers will cover a wavelength range from 0.3\,mm to 9\,mm. The
intensity at the different wavelengths originates from different
layers from the high photosphere to the middle chromosphere, with the
average formation height increasing with wavelength.  Simultaneous
observation at different wavelengths could thus serve as a tomography
of the atmosphere, finally allowing for mapping the
(three-dimensional) thermal structure of the solar chromosphere
\cite[(Loukitcheva \etal\ 2006; Wedemeyer-B\"ohm \etal\ in prep.)]{loukitcheva06,wedemeyer_alma}. 
The intensity contrast at 1\,mm is 23\,\% and thus very similar to
the 500\,nm continuum. 

\section{Discussion and Conclusions}\label{sec:concl}

Despite the steady increase in realism many details of the models
still have to be taken with caution. Maybe the strongest limitation is
the assumption of LTE (in particular for the radiative transfer) which
was necessary to keep the problem computationally tractable so far.
LTE is also assumed in the ``post-simulation'' radiative transfer
calculations with \textsf{Linfor3D}.  The extension to a more
realistic non-LTE treatment in both the simulation and the
intensity/spectrum synthesis will be an important point to work on in
future.  Consequently, the temperature amplitudes of the current model
chromosphere might still be uncertain to some extent.

Nevertheless this class of models can give important hints for the
interpretation of observations.  Co-existing cool and hot regions in an
inhomogeneous and dynamic atmosphere offer the ultimate solution for
the diagnostic dilemma with apparently contradicting temperatures
derived from CO on the one and UV-based diagnostics on the other hand.
These different observational aspects can now be explained together with one 
single model, like the one presented here, whereas conventional 
(semi)-empirical 1D models 
\cite[(e.g., Vernazza \etal\ 1981)]{val81}
can at best account for selected aspects of the atmosphere.  The full
potential of detailed 3D models gets obvious when addressing the
small-scale spatial pattern.  For instance, the models imply that the
spatial resolution of TRACE at 160\,nm is not sufficient to resolve
the small-scale pattern of the middle to upper photosphere. 
That has consequences for the conclusions derived for atmospheric 
dynamics and heating 
\cite[(Fossum \& Carlsson 2005)]{2005Natur.435..919F}.

A comprehensive numerical model is also helpful for defining
constraints on diagnostics (in combination used as a kind of tomography)
that are required for the desired empirical 
determination of the thermal structure of the atmosphere.  
For instance, the continua at (sub-)millimetre
wavelengths as chromospheric diagnostics seem to be a promising
alternative to the complicated inner regions of the Ca\,{II} resonance
lines.

\begin{acknowledgments}
  I would like to thank my colleagues B.~Freytag, M.~Steffen,
  H.-G.~Ludwig, I.~Kamp, J.~Leenaarts, O.~Steiner, J.~Bruls, and 
  R.~Schlichenmaier for
  their contributions to various related projects and for many
  inspiring discussions. SW was supported by the {\em Deutsche
    Forschungs\-gemein\-schaft (DFG)}, project Ste~615/5.
\end{acknowledgments}


\begin{discussion}

  \discuss{Kupka}{Have you looked at the effects of different upper
    boundary conditions on your simulations?}

  \discuss{Wedemeyer-B\"ohm}{Yes. We use a transmitting upper boundary
    and checked its influence for the hydrodynamics case and found no
    (artificial) reflection of waves (as it would be the case for a
    closed boundary). The further development of the boundary
    conditions for the MHD case is in progress but the current version
    already seems to work properly.  \textit{There will be a new open
      lower and a transmitting upper MHD boundary condition available
      soon which will be checked thoroughly.}}

  \discuss{Martinez Pillet}{What happens with the CO clouds, if you
    start with mean fields larger than 10\,G ?}

  \discuss{Wedemeyer-B\"ohm}{We have carried out 2D simulations with
    different field strengths which show a clear difference in the
    distribution of molecules \textit{(Wedemeyer-B\"ohm \etal\ 2005,
      ESA SP-596, 117)}.  For a systematic study, however, we would
    like to compute a sequence of 3D models first.}

\end{discussion}

\end{document}